\documentstyle[twoside,fleqn,npb,epsfig]{article}

\title{DIANA and selected applications\thanks{Supported by EU project
HPRN-CT-2000-00149}}
\def\theaddress{\relax}
\author{J.~Fleischer\address{Universit\"at Bielefeld,
        Fakult\"at f\"ur Physik, Bielefeld, Germany},%
        \addtocounter{address}{-1}
        M.~Tentyukov\addressmark\thanks{Supported by DFG under the project
FL 241/4-2}\thanks{On leave from JINR, Dubna, Russia}
        and\addtocounter{address}{-1}
        O.V.~Tarasov\addressmark\footnotemark[2]\footnotemark[3]
        }
\begin{document}

\begin{abstract}
New developments concerning the extension of the Feynman diagram analyzer
DIANA are presented. We discuss new graphics facilities, different approaches
to automation of momenta distribution and parallel processing facilities.
Furthermore applications to $t\bar t$ production and Bhabha scattering are
shortly discussed.
%\vspace{1pc}
\end{abstract}

\maketitle
\setcounter{footnote}{0}
The project called DIANA (DIagram ANAlyzer)\cite{Diana} for the evaluation
of Feynman diagrams was started by our group some time ago. It was already
used to calculate several processes \cite{usage}. The recent
development\footnote{For details look at\\
http://www.physik.uni-bielefeld.de/$\tilde{\,}$tentukov/diana.html}
of this project is do\-cumented in this contribution.

The {\bf pictorial representation} of diagrams described in \cite{graf}
includes three different kinds of postscript files. Now one more
kind is available, the encapsulated postscript file containing
particle lines together with momenta flow
\cite{dianacat02}\footnote{
For details see also http://www.physik.uni-bielefeld.de/\par
{\hfill \~{}tentukov/printing.html}}:

\vskip 2mm
%\begin{figure}[ht]
%\vskip 80mm
\centerline{\epsfxsize=45mm \epsfbox{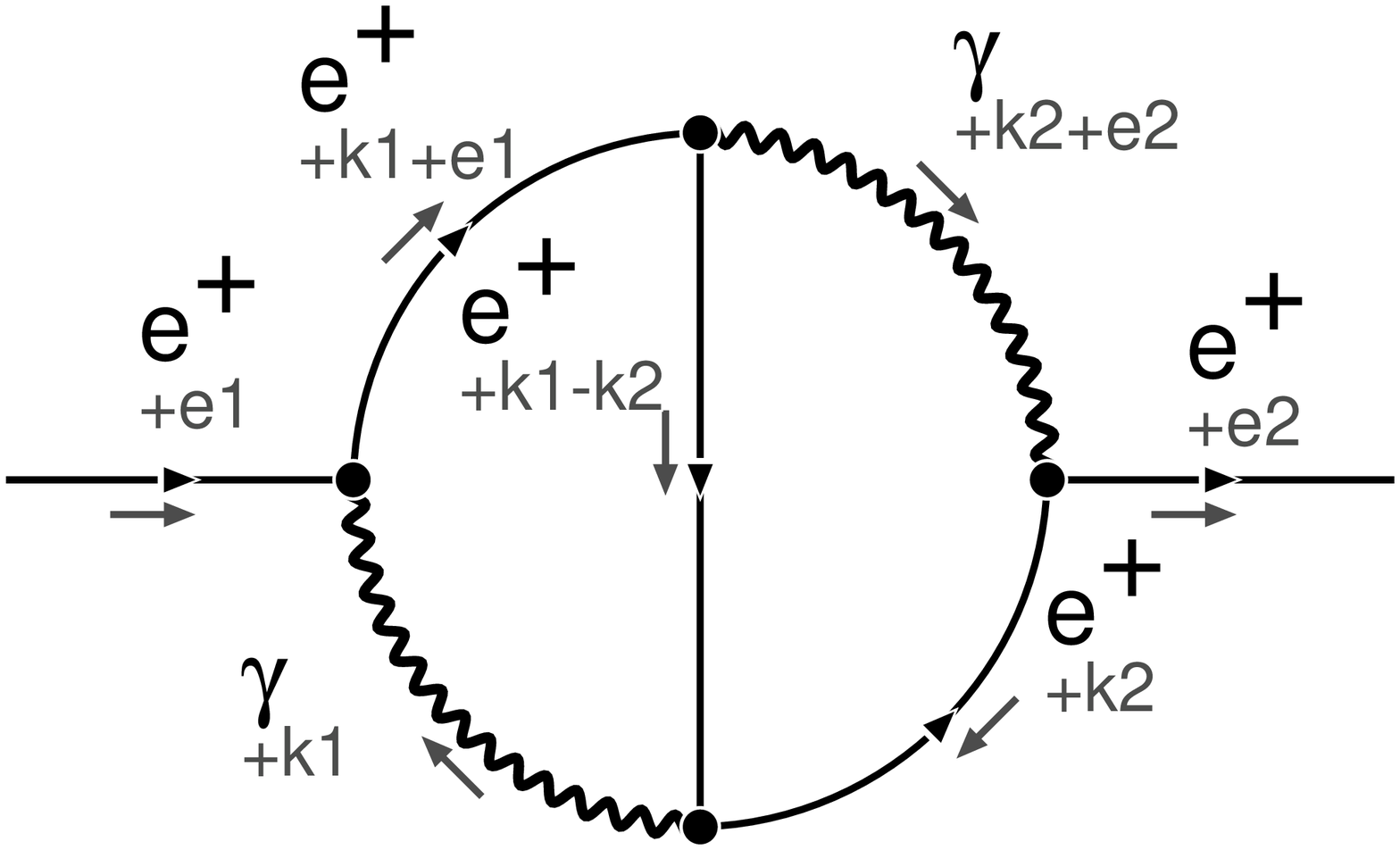}}
%\caption{
%   \label{dd1}
%        }
%\end{figure}

\vskip 1mm

To use it, the user
invokes the function \\
\verb|\outInfoEPS(Fname,Height,Font,Fontsize)|.\\
 The picture will be saved in the file
 \verb|Fname|.
       If \verb|Fontsize| = 0 the particle
images (e.g. $e^\pm$) will not be printed. The width of the
       diagram will be calculated from ``\verb|Height|''. The diagram will be scaled to
fit the EPS bounding box [0 0 \verb|Width| \verb|Height|].

Some parameters may be adjusted by the user. There are two functions,
\verb|\setIEPSshift()|
and
       \verb|\setIEPSpars()|, which may be invoked in the ``initialization''
section to reset default parameters, see\\
{\footnotesize
http://www.physik.uni-bielefeld.de/\par\vspace*{-1mm}
{\hfill \~{}tentukov/printing.html\#OUTINFOEPS}
}

Particle identifiers can now be
depicted by different fonts, sizes and colours. By default the particle image will
be printed as in the corresponding entry
 of the propagator specified in the `model file'. The user can set
other images by means of
the model extension, added in version 2.28, or by using
the function   \verb|\setparticleimage(A,image)|.
The particle images are produced by specific constructions, e.g.:
\verb|W{y(10)+}| $\to$ ${\sf W^+}$.

`Blocks' allow local shifting etc. of certain characters.
A block is started by `\{` followed
by a keyword with parameters
    in parenthesis:

        \verb|{x(#) ...  }| - paints the content shifted along the abscissa
(x). After the
block, the current point is set to (old x, new y);

    \verb|{y(#) ...  }| - paints the content shifted along the ordinate (y).
    After the block,
the current point is set to (new x, old y);

    \verb|{xy(#)(#) ...  }| - paints the content with shifts along both x and
y. After the block, the current point is set to (old
    x, old y);

    \verb|{f(fontname)(#) ...  }| - sets font \verb|fontname| scaled by \# (in
 fractions of 1/10 of the current font size);

    \verb|{s(#) ...  }| - scales the current font by \# (in
    fractions of 1/10 of the
current font size);

    \verb|{c(#)(#)(#) ...  }| - sets RGB\footnote{Red-Green-Blue -- one of a
standard colour model.} colour. Each parameter must be a
positive decimal number in the region $[0{\ldots} 1]$.

    All sizes are in fractions of 1/10 of the base font size, i.e. ``10'' means
the size of the current font, ``20'' is twice bigger,
    ``5'' is  half size.

        As \verb|fontname| it is recommended to use only the standard 13
PostScript level 1 fonts:

\centerline{\epsfxsize=\linewidth \epsfbox{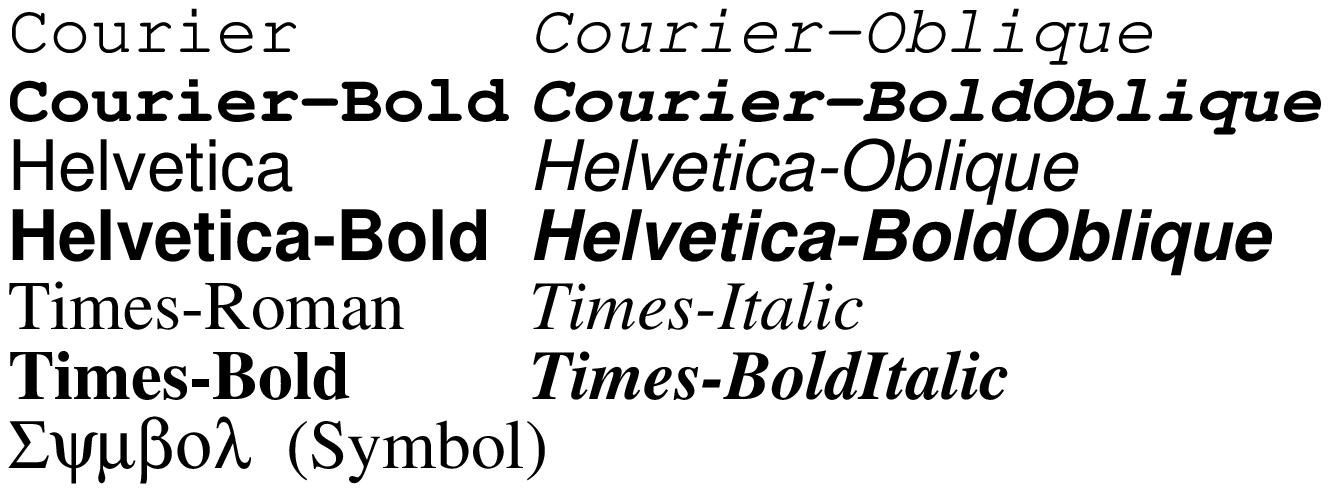}}

To set the image in the model, the user
must add, after a line type, the character `;' and the particle image.
  Example:\\
\vspace*{-6mm}
\begin{verbatim}
[Wm,Wp;W;VP(num,ind:2,ind:1,vec,2);mmW;
        arrowWavy,3,2;W{y(10)-}]
\end{verbatim}
\vspace*{-2mm}
    In this case the line will be of a photon-like arrowed curve of  width 2 and
amplitude 3, and the particle image will be ${\sf W^-}$.

As alternative, the function\\ \verb|\setparticleimage(particle,image)|
overwrites the image definition coming from the model.

\bigskip

Topologies are represented in \linebreak terms of
    ordered pairs of numbers like \linebreak \verb|(fromvertex,
tovertex)|\cite{Diana}.
    All external legs have negative numbers.

Often the number of topologies is too large such that
it is impossible to {\bf assign
momenta} to the lines in all topologies ``by hand''.
In that case, without special action,  momenta are  introduced automatically
in an arbitrary manner (of course taking care of momenta conservation).
The user names the loop
momenta via the macro \verb|\loopmomenta| in the ``create'' file, e.g.
\verb|\loopmomenta(k1,k2,k3)|,
and DIANA will assign momenta automatically using ``$k1$'', ``$k2$'' and
``$k3$'' as the
loop integration momenta.

Sometimes it is important to keep some definite lines free from
external momenta.
If the users specifies \verb|SET _MARK_LOOP=YES| in the
``create'' file, the topology editor will be invoked in a special mode, and
the user ``clicks'' (by mouse) which lines should carry bare
integration momenta. All remaining momenta will be assigned automatically.

Sometimes it is
necessary to use more sophisticated distributions.
For example, the user may
want to use his favorite momenta like $k-p_1, k-p_2, k-p_3$, etc. assigned to
some definite lines (see e.g. \cite{FTJ9907327}). For such cases, DIANA
provides the possibility to define momenta only for the virtual lines, and
the full set of topologies (obtained by exchanging external momenta)
will be defined from these ``internal'' topologies
attaching the external legs. In this case momenta for
internal parts are defined in terms of combinations of loop momenta ($k$ in
the above example) and some ``abstract'' tokens ($p_1, p_2$ etc.), and for
each topology DIANA expresses these tokens in terms of external momenta.

Another example: occasionally topologies are generated by more complicated
(``generic'') ones by scratching lines. In such cases one wants to stick to
the momenta introduced for the lines which are kept \cite{dianacat02}\footnote{
\parbox[t]{\linewidth}{For details look at http://www.physik.uni-bielefeld.de/\par
{\hfill \~{}tentukov/generictopology.html}}}.
E.g. the user
investigates the generic topology\\
\vspace*{-7mm}
\begin{verbatim}
generictopology A =
 (-2,2)(-1,1)(1,3)(3,2)(2,4)(4,1)(3,4):
 p1+k2,p1+k1,k1,k2,k2-k1.
\end{verbatim}
\vspace*{-2mm}
Then DIANA will generate topologies from this one
by scratching lines in the following manner:

\vskip 2mm
\centerline{\epsfxsize=50mm \epsfbox{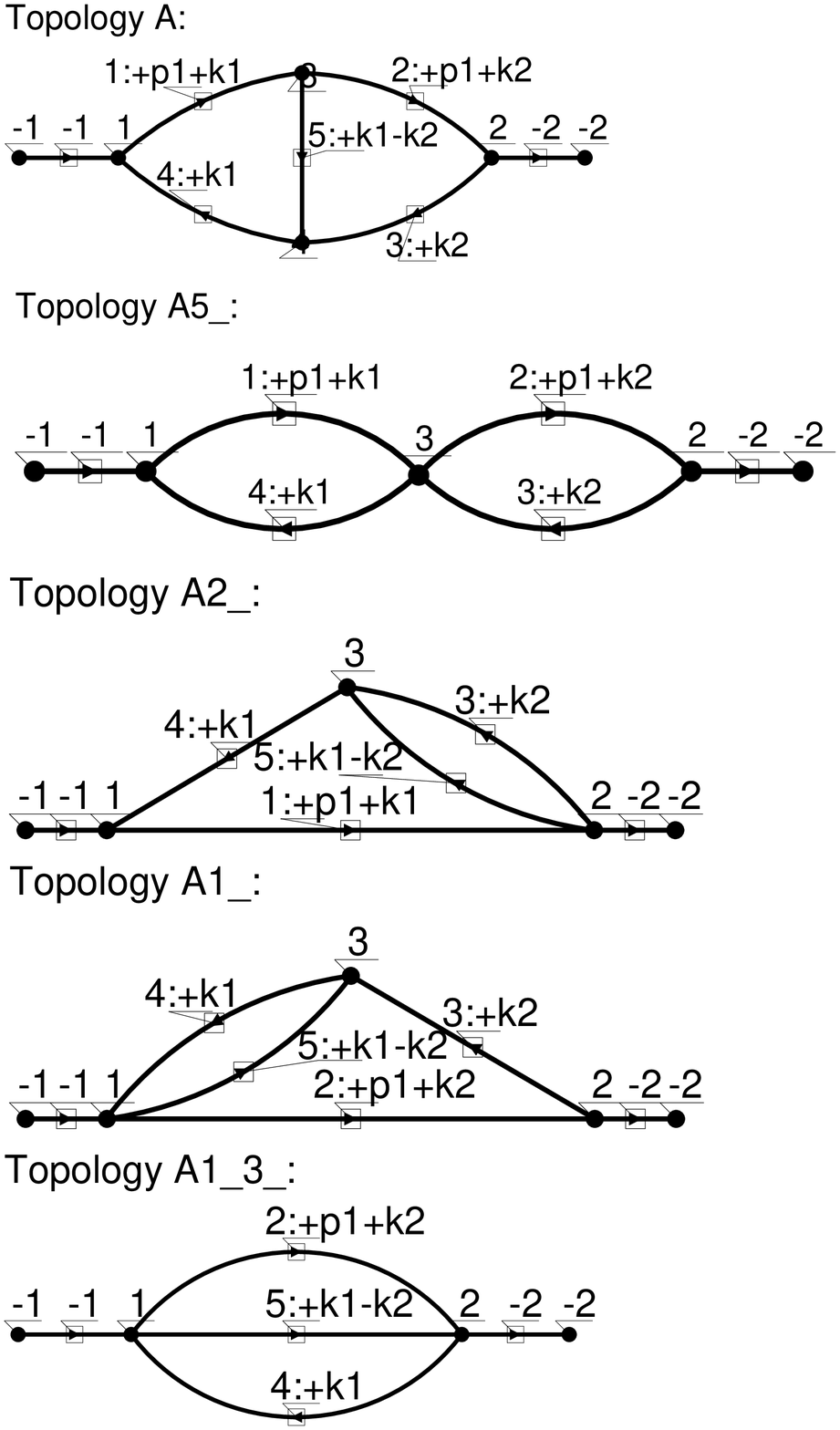}}
%\vskip 2mm

The topology name is constructed by DIANA as follows: the name of the
``generic'' topology is appended by the index of ``scratched'' lines separated
by underscore ``\_''.
%%%%%%%%%%%%%%%%%%%%%%%%%%%%%%%%%%%

\bigskip

To run an external command \verb|cmd|, the function \verb|\exec(cmd)| can be
used. This function exe\-cutes the command in the background without waiting
for it to be completed. So, this function may be used to {\bf paralyse} the
evaluation of a process by running more then one FORM job simultaneously.

To avoid overloading the processor, by default at each time only one job
is actually running while all the rest is waiting in the queue.
On SMP\footnote{Symmetric MultiProcessing} computers, the number of
simultaneously running jobs can by
changed by means of a command line option \verb|-smp|. Thus, \verb|-smp 8|
tells DIANA to run on the computer 8 jobs simultaneously while all the rest
is queued.

To synchronize the TM\footnote{TM is an abbreviation for ``Text Manipulating'', see
\cite{Diana}.} program with all started jobs, the function
\verb|\waitall(timeout)| is used. It suspends execution of the TM program
until all jobs are completed. Each ``\verb|timeout|'' milliseconds the function
reports the number of jobs which are not finished yet.

In case of a cluster of computers with a current directory shared by means of
NFS\footnote{Network File System} e.g., the function \verb|\exec()| can use
DIANA servers running on other computers.
To run DIANA as a server, the user
must use the command line option \verb|-d #|, where \verb|#| is the number
of jobs which can run on this computer simultaneously. Thus, each computer on
which the user has executed the command
\verb|diana -d 1 -q|
is ready to perform the commands queued by the function
\verb|\exec()| (we assume that the current directory is shared by NFS among
all computers in the cluster). Here \verb|-d 1| tells DIANA to run a
daemon\footnote{A program running in the background and listening to some port.}
accepting only one connection, and the option \verb|-q| terminates the father
DIANA process. For SMP computers the optimal argument for the \verb|-d| option would
be the number of processors.

Two functions, \verb|\exec()| and \verb|\waitall()|, permit the user to
organize a simple parallel session for evaluations on SMP
and/or cluster of independent computers with shared disk space. But, very
often this is not enough. Indeed, let us suppose that all results must be
collected into one resulting file \verb|log.all|, while every job produces a
file \verb|log.#| where \verb|#| is the order number of the job. Since jobs
are running and completing independently from each other, we can only
collect  all \verb|log.#| into \verb|log.all| after {\em all} the jobs are
finished. This leads to producing a lot of files \verb|log.#| at
an intermediate step, which can overload a file system.

The simple solution is to allow some of newly started jobs to be synchronized
with all previously started jobs. Indeed, in this case after each job we can
start another ``slave'' job, which appends the \verb|log.#| file containing the
result of the ``master'' job to the file \verb|log.all|. To obtain a correct
order of the file \verb|log.all|, this copying job must be performed only
after all earlier jobs are finished.

Another problem with cluster computations is that the optimal placement of
the resulting file \verb|log.#| is usually a \verb|/tmp| directory which is
local against a current node\footnote{DIANA
assumes the ``node'' is the IP address of a server, so the conception of
``nodes'' is actually supported only for clusters. For SMP, the whole computer
is assumed to be a single node.}, on which a job ``number \verb|#|'' is performed.
But to do this, the ``slave'' copying job has to ``stick'' to the ``master'' job,
i.e.  it must be performed on the same node as the "master" job.

In order to create such a ``slave'' job, the function \verb|\stick(cmd)|
must be used.
This function is similar to  \verb|\exec(cmd)| with two
exceptions.
It
performs the command \verb|cmd| only after all earlier jobs are completed.
This function in general is used to sum up all produced files.
That is why it performs \verb|cmd| on the same computer as the preceding job.

As an example, let us consider the following script \verb|runf|:
{\footnotesize
%\begin{figure*}[t]
\begin{verbatim}
#!diana -smp 1 -c runpar.tml
\STARTSERVERS(phya25,phya26,phya27,phya28,phya29)
\system(echo > log.all)

\REPEAT(N)
\exec(form -d i=\get(N) do.frm > /tmp/log.\get(N))
\stick(cat /tmp/log.\get(N) >> log.all)
\stick(rm /tmp/log.\get(N))
\ENDREPEAT()

\waitall(2000)
\end{verbatim}
%\caption{\label{thescript}}
%\end{figure*}
}
\noindent
The user enters: \verb|runf 186 200 |, and the system executes:\\
\verb|diana -smp 1 -c runpar.tml runf 186 200|.

The file \verb|runpar.tml| contains definitions of various TM functions and
other settings; in particular, it redefines the comment character as
\verb|#|. The macro \verb|\STARTSERVERS(list)| checks if each server is working
and, if not, it starts the server by means of the \verb|ssh|. For example, for
the host phya26 the following command will be performed:
\begin{verbatim}
ssh phya26 cd CD ; diana -d 1 -q
\end{verbatim}
where ``\verb|CD|'' is a current directory, e.g. \verb|/home/user/jobs|.

The operator \verb|\system(cmd)| executes the command \verb|cmd| synchronously,
i.e. it waits for the command to be completed and returns an exit code. Here it
is used to produce an empty file ``\verb|log.all|''.

All the instructions between \linebreak \verb|\REPEAT(N)...\ENDREPEAT()| are cycled
with \verb|N=186,...,200|.
We assume that there is some folder file, say, \verb|tt.in| with FORM input
for each diagram. The FORM program \verb|do.frm| evaluates a
diagram by virtue of including a fold from the folder \verb|tt.in| via
an instruction like
\verb|#include tt.in # n'i'|. The macro definition \verb|i| comes from
the command line \verb|form -d i=\get(N) do.frm|, where \verb|\get(N)|
runs from 186 to 200. Each FORM job saves the result to the local
directory, but the corresponding concatenation is performed by \verb|\stick(cat ...)|
on the same computer. At the end, all results will be collected in the file
\verb|log.all|, and all intermediate files \verb|\tmp\log.#| will be removed.

After all jobs are queued, the function \verb|\waitall(2000)| will report
every 2 seconds how many jobs are not yet completed.\\

%\cite{FTJ9907327}

{\bf Recent applications} of DIANA in our collaboration are $e^+ e^-$ annihilation
into $t \bar{t}$ and Bhabha scattering.\\

The $t \bar{t}$ production has been calculated including hard bremsstrahlung
and various comparisons with other groups have been performed. For further
information about the results we refer to recent notes on this subject
\cite{tbart1,tbart2}.\\

Concerning Bhabha scattering, since we are heading for two-loop calculations,
we perform all calculations in arbitrary dimension $d=4-2 \varepsilon$. The
algebraic part of this calculation has been presented in \cite{Bhab}. Here we
discuss only the eva\-luation of one-loop master integrals,
in particular of box diagrams. The method applied uses eva\-luation of difference
equations \cite{FJT}. We investigate now the following (scalar) box diagram with
two (zero mass) photons in the s-channel:

\vskip 1mm
\centerline{\epsfxsize=45mm \epsfbox{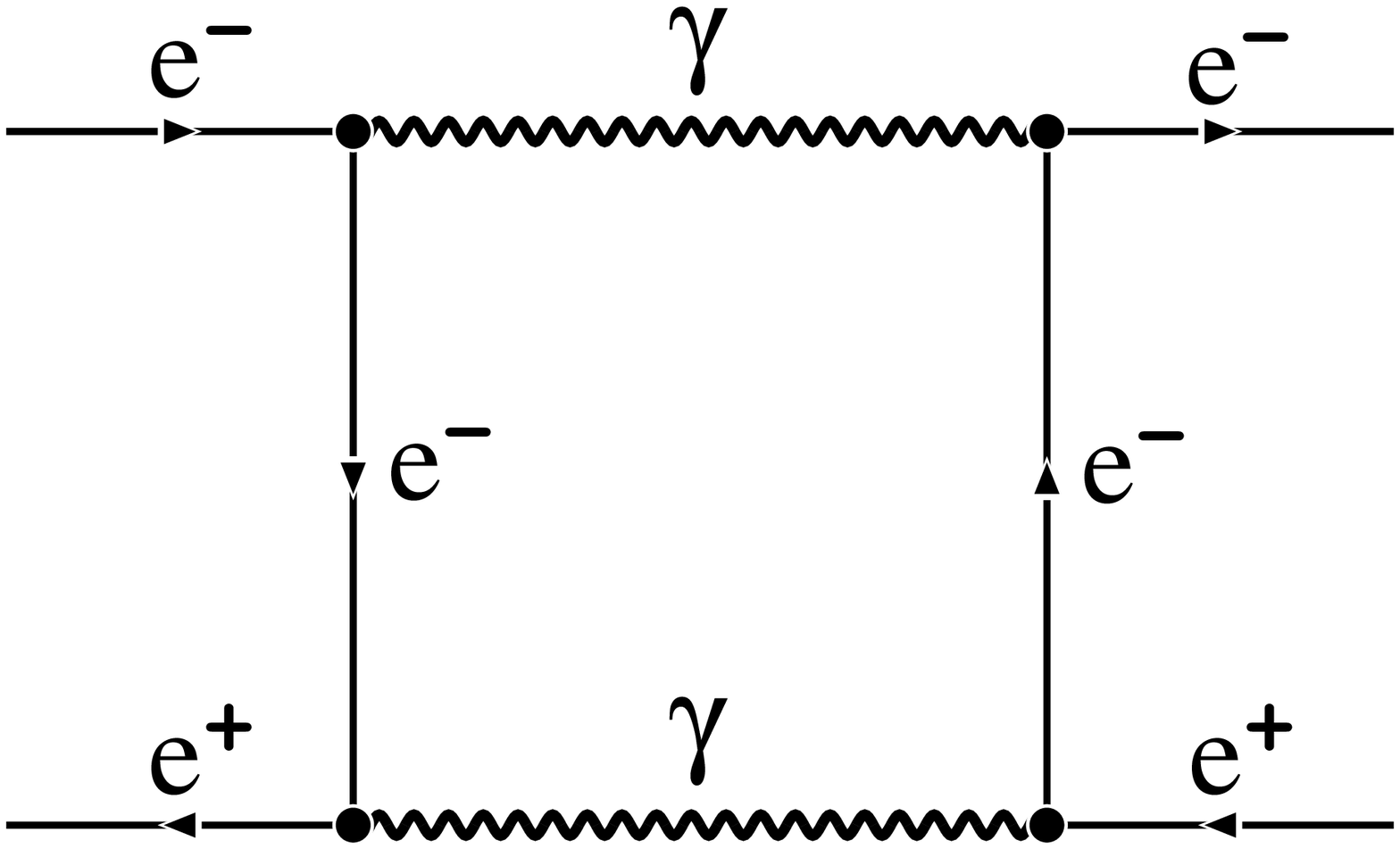}}
\vskip -7mm

\vspace{0.5cm}

The corresponding master integral in d dimensions can be represented as
\begin{eqnarray}
&&\hspace{-0.6cm} I_4^{(d)} = \nonumber \\
&&\hspace{-0.6cm} \frac{2 \Gamma(2-\frac{d}{2})}{t-4 m^2} \biggl[ \nonumber \\
&&\hspace{-0.6cm} \frac{(m^2)^{(\frac{d}{2}-2)}}{s} \int_0^1{\frac{dx~ x^{\frac{d-5}{2}}}{1-k_1 x}}
                          {~}_2F_1(1,\frac{d-3}{2},\frac{d-2}{2},k_2 x) \nonumber \\
&&\hspace{-0.6cm} -\frac{ \sqrt{\pi} (-s)^{(\frac{d}{2}-2)} } {m \sqrt{s}~ 2^{(d-3)} } ~
    \frac{\Gamma \left(\frac{d-2}{2} \right)  } {\Gamma \left(\frac{d-3}{2}\right) }~
    \int_0^1{\frac{dx~ x^{\frac{d-5}{2}}}{1-k_3 x}~ \frac{1} {\sqrt{1- k_4 x}}} \nonumber \\
&&\hspace{-0.6cm} -\frac{(m^2)^{(\frac{d}{2}-2)}}{s} \frac{\Gamma \left(\frac{d-2}{2} \right)  } {\Gamma
    \left(\frac{d-3}{2}\right) }~ \sqrt{1-\frac{4 m^2}{t}}~\biggl( \nonumber \\
&&\hspace{-0.6cm} (1-\frac{t}{4 m^2})^{(\frac{d}{2}-2)}  \sqrt{\pi}~ \Phi(k_5,1,\frac{d}{2}-2) \nonumber \\
&&\hspace{-0.6cm} -(1-\frac{4 m^2}{t})^{(\frac{d}{2}-2)} \frac{1}{\sqrt{\pi}}
    \int_0^1{\frac{dx~ x^{\frac{d-5}{2}}}{\sqrt{1- x}}} (1-k_6 x)^{(2-\frac{d}{2})} \nonumber \\
&&\hspace{-0.6cm}  \Phi(\frac{-k_5 k_6 x}{1-k_6 x},1,\frac{d}{2}-2)  \biggr)
    \biggr], \nonumber \\
\nonumber
\end{eqnarray}
where the $k_i ( i=1 \dots 6 )$ are the following kinematical variables
%\newpage
\begin{eqnarray}
&&k_1 = 1 - \frac{4 m^2}{s}~~~~~,~~~   \nonumber \\
&&k_2 = -4 m^2 (\frac{1}{s}+\frac{1}{t-4 m^2}) \equiv - m^2 z \nonumber \\
&&k_3 = 1 + \frac{s}{t-4 m^2},~~~k_4 = 1 - \frac{s}{4 m^2}  \nonumber \\
&&k_5 = 1 + \frac{t-4 m^2}{s},~~~k_6 = \frac{4 m^2}{t} \nonumber \\
\nonumber
\end{eqnarray}
and $\Phi(z,1,a)$ is the Lerch function
$$
\Phi(z,1,a)=\sum_{k=0}^{\infty} {\frac{z^k}{k+a}} \nonumber \\
$$
   For $s > 4 m^2$ and $t < 0$ it is easy to see that we remain in the analyticity
domain of all occurring expressions, except for the factor $(-s)^{(\frac{d}{2}-2)}$.
This result finally allows to expand the diagram in $\varepsilon = \frac{4-d}{2}$.
In the two-loop calculation we need to expand up to order $O(\varepsilon)$, which
will yield finite contributions when multiplied, e.g., by divergent counterterms,
containing terms of order $O(\frac{1}{\varepsilon})$. Due to the overall factor
$\Gamma(2-\frac{d}{2})$, coming from the infrared divergence of the diagram,
the expansion of the various contributions in square
brackets must be performed up to order $O(\varepsilon^2)$.
We also see that the imaginary part of the finite part of the diagram comes from
the above  $(-s)^{(\frac{d}{2}-2)}$ when expanded up to order $O(\varepsilon)$.\\

   Crossed diagrams can be obtained from this form as well if the result of
the expansion yields analytic expressions (for which the analytic continuation
should be known in general) or again the kinematical variables are such that a
nume\-rical integration hits no singularities.\\

   {\bf Acknowledgement} The authors wish to thank the organizers of
`RADCOR' and `Loops and Legs 2002' for the nice organization of the conference.
M.T. and O.V.T thank the DFG for financial support. J.F. and O.V.T want to thank
DESY Zeuthen for travel support for visits.

\end{document}